\documentclass[aps,twocolumn,nofootinbib,preprintnumbers,superscriptaddress]{revtex4}
\usepackage{hyperref}
\usepackage{graphicx}
\newcommand{\bea}{\begin{eqnarray}}
\newcommand{\eea}{\end{eqnarray}}

\newcommand{\be}{\begin{equation}}
\newcommand{\ee}{\end{equation}}
\newcommand{\beas}{\begin{eqnarray*}}
\newcommand{\eeas}{\end{eqnarray*}}

\def\Biggg#1{{\hbox{$\left#1\vbox to 25pt{}\right.\n@space$}}}
\def\n@space{\nulldelimiterspace=0pt \m@th}
\def\m@th{\mathsurround = 0pt}

\begin{document}

\preprint{YITP-13-73}

\title{On the Mutual Information in Hawking Radiation}

\author{Norihiro Iizuka}
\email[]{iizuka@yukawa.kyoto-u.ac.jp} 
\affiliation{Yukawa Institute for Theoretical Physics, 
Kyoto University, Kyoto 606-8502, JAPAN}

\author{Daniel Kabat}
\email[]{daniel.kabat@lehman.cuny.edu}
\affiliation{Department of Physics and Astronomy,
Lehman College, City University of New York, Bronx NY 10468, USA}

\begin{abstract}
\noindent
We compute the mutual information of two Hawking particles emitted
consecutively by an evaporating black hole.  Following Page, we find
that the mutual information is of order $e^{-S}$ where $S$ is the
entropy of the black hole.  We speculate on implications for black
hole unitarity, in particular on a possible failure of locality at
large distances.
\end{abstract}


\maketitle

\noindent
Hawking's discovery that black holes emit thermal radiation
\cite{Hawking:1974sw} is one of the few tangible results in quantum
gravity, and the resulting conflict with unitarity
\cite{Hawking:1976ra} has driven much of the research in the field.
See \cite{Mathur:2009hf} for a review.  The goal of the present paper
is to obtain new insight into this issue, from a computation of the
mutual information carried by successive Hawking particles.

\begin{figure}[h]
\begin{center}
\includegraphics{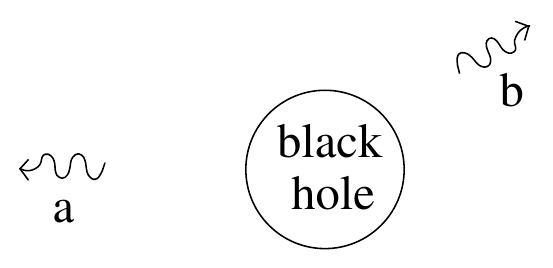}
\end{center}
\caption{Successive Hawking particles emitted by a black hole.\label{figure}}
\end{figure}

Consider two successive Hawking particles emitted by an evaporating
black hole, as shown in Fig.\ \ref{figure}.  Motivated by the AdS/CFT
correspondence we assume that a conventional quantum mechanical
description of this process is available.  In particular we assume
there is an underlying Hilbert space with unitary time evolution that
describes the microscopic degrees of freedom.  In this fine-grained
description there is no tension with unitarity: the two Hawking
particles are correlated due to their shared history, in which they
both interacted with the microscopic black hole degrees of freedom.
In this fine-grained description, Hawking radiation from a black hole
is no different from the blackbody radiation emitted by any other hot
macroscopic object.

However for a black hole we would like to consider a coarse-grained
description, in which the black hole is characterized just by its
macroscopic thermodynamic properties such as energy and and entropy.
It seems reasonable that this coarse-graining gives rise to the usual
notion of a semiclassical spacetime.  That is, only in a
coarse-grained description could one hope to describe the black hole
using the usual Schwarzschild metric, and could one hope to describe
Hawking radiation using effective field theory on the Schwarzschild
background.  In support of this view, note that the usual black hole
metric only captures macroscopic properties such as mass or charge.
Also Hawking's calculation, carried out in this context, shows that a
black hole emits uncorrelated thermal radiation.  This behavior is
expected in a coarse-grained description of blackbody radiation, since
such radiation is completely characterized by a macroscopic quantity,
namely the temperature of the black hole.

In this setting, to understand unitarity, the main challenge is
identifying which properties of the coarse-grained description deviate
most significantly from the underlying microscopic description.  Given
some underlying microscopic theory, what modification to the
coarse-grained description is most appropriate for restoring
unitarity?

To sharpen our discussion we consider the correlation between the two
successive Hawking particles $a$ and $b$ shown in Fig.\ \ref{figure}.
We have in mind two particles that are emitted almost simultaneously
from well-separated points on the horizon, so that the separation
between $a$ and $b$ is large and spacelike.  The correlation can be
measured by the mutual information
\be
\label{Iab}
I_{ab} = S_a + S_b - S_{ab}
\ee
where $S_a$ is the entropy of particle $a$, $S_b$ is the entropy of
particle $b$, and $S_{ab}$ is the entropy of both.  According to
Hawking's calculation $a$ and $b$ are uncorrelated and the mutual
information vanishes.  Under seemingly reasonable assumptions this
will remain true even in the presence of interactions
\cite{Mathur:2009hf}.  But if the entire system (including the black
hole) is in a random pure state, the true correlation between $a$ and
$b$ can be obtained from the fundamental work of Page
\cite{Page:1993df}.  Page considers a Hilbert space of dimension $m$
entangled with another Hilbert space of dimension $n \geq m$, and shows
that in a random pure state the average entropy is
\be
\label{Page}
S_{m,n} = \sum_{k = n+1}^{mn} {1 \over k} - {m - 1 \over 2 n}
\ee
For large $n$ the sum can be estimated using the Euler-Maclaurin
formula, which gives
\be
\label{Page2}
S_{m,n} = \log m - {m^2 - 1 \over 2 m n} + {\cal O}(1/n^2)
\ee
To apply this to the situation at hand, let $N_a$ be the dimension of
the Hilbert space of particle $a$, let $N_b$ be the dimension of the
Hilbert space of particle $b$, and let $N_{bh} = e^S$ be the dimension of
the Hilbert space of the black hole.  For particle $a$, for example,
we have a Hilbert space of dimension $N_a$ entangled with a Hilbert
space of dimension $N_b N_{bh}$.  Thus
\bea
\nonumber
 S_a &=& S_{N_a,N_bN_{bh}} \\
 S_b &=& S_{N_b,N_aN_{bh}} \\
\nonumber
 S_{ab} &=& S_{N_aN_b,N_{bh}}
\eea
Using (\ref{Iab}) and (\ref{Page2}) we find that for large $N_{bh}$, the
mutual information in the Hawking particles $a$ and $b$ is
\bea
\label{MutualInformation}
I_{ab}&=& {(N_a^2 - 1)(N_b^2 - 1) \over 2 N_a N_b N_{bh}} + {\cal O}(1/N_{bh}^2)
\eea
This is our main result.  It shows that the mutual information carried
by two successive Hawking particles is of order $e^{-S}$.  For example
if each Hawking particle could carry one bit of information then $N_a
= N_b = 2$ and $I_{ab} \approx {9 \over 8} e^{-S}$, while if each
Hawking particle could carry a large amount of information then
$I_{ab} \approx {1 \over 2} N_a N_b e^{-S}$.

As we discussed above, the usual semiclassical picture of gravity must
be modified in order to reproduce these correlations.  Roughly
speaking the possible modifications fall into three categories.
\begin{enumerate}
\item
Modify the interior \\[5pt]
It could be that microscopic quantum gravity effects become important
at or inside the (stretched) horizon of the black hole, invalidating
the use of the classical Schwarzschild geometry in this region and 
generating correlations between outgoing Hawking particles.
However outside the horizon semiclassical gravity and effective field
theory could be valid.  Proposals of this type include fuzzballs
\cite{Mathur:2005zp,Mathur:2008nj} and firewalls
\cite{Almheiri:2012rt,Almheiri:2013hfa}. 
\item
Modify the exterior \\[5pt]
It could be that effective field theory is not trustworthy, even at
macroscopic distances outside the black hole.  For example, it could
be that the underlying theory of quantum gravity leads to violations
of locality over large distances, in a way that generates correlations
and restores unitarity.  Some models with non-locality have been
discussed in \cite{Giddings:2011ks,Giddings:2012gc,Giddings:2013kcj}.
\item
Modify both \\[5pt]
Perhaps both the interior region of the black hole and the rules of
effective field theory outside the black hole receive important
corrections due to microscopic quantum gravity effects.
\end{enumerate}
Unfortunately, just from considerations of unitarity, there is no
clear way to decide between these possibilities.  But since most
models discussed in the literature take other approaches, let
us indulge in a little speculation about the
possibility of non-locality outside the horizon.

A key principle in local field theory is microcausality, that is, the
property that field operators commute at spacelike separation.  If we
are prepared to give up on locality outside the horizon, it could
be that spacelike separated field operators no longer commute.  We have
in mind that the resulting non-locality extends over macroscopic
distances, and would thus fall into the category of modifying the exterior
of the black hole.  But we must admit that in order to restore unitarity,
non-locality which extends to the stretched horizon could do the
job.

In fact, AdS/CFT may provide some motivation for the radical idea of
non-locality over macroscopic distances.  Order-by-order in the $1/N$
expansion of the CFT one can construct CFT operators which mimic local
field operators in the bulk \cite{Kabat:2011rz,Heemskerk:2012mn}.  The
algorithm involves starting from a single primary field and adding an
infinite tower of higher dimension operators.  In the $1/N$ expansion
one can show that the resulting CFT operators commute whenever the
bulk points are spacelike separated.\footnote{In fact bulk
microcausality can be taken as a guiding principle for the
construction of bulk observables \cite{Kabat:2011rz}.  The procedure
is simplest for bulk scalars, but it works for gauge fields as well
\cite{Kabat:2012av}.}  But at finite $N$ it seems unlikely that the
higher dimension operators required for bulk locality could exist.
Instead it's more likely that bulk observables will fail to commute at
spacelike separation, even over macroscopic distances, by an amount
which is non-perturbatively small in the $1/N$ expansion.

As a toy model for this idea, consider a pair of independent harmonic oscillators characterized by
\be
[\hat{\alpha},\hat{\alpha}^\dagger] = [\hat{\beta},\hat{\beta}^\dagger] = 1
\ee
with all other commutators vanishing.  We think of these oscillators
as representing two independent degrees of freedom in some underlying
microscopic description (``the boundary'').  Suppose these boundary
operators can be mapped to bulk operators which describe the Hawking
particles shown in Fig.\ \ref{figure}.  We assume a boundary-to-bulk
map depending on two parameters $\theta$ and $\phi$, explicitly given
by
\bea
\label{Bog1}
&& \hat{a} = \hat{\alpha} \cosh(\theta + \phi) + \hat{\beta}^\dagger \sinh(\theta + \phi) \\
\label{Bog2}
&& \hat{b} = \hat{\beta} \cosh(\theta - \phi) + \hat{\alpha}^\dagger \sinh(\theta - \phi) \\[5pt]
\label{Bog3}
&& \hat{a}^\dagger = \hat{\alpha}^\dagger \cosh(\theta + \phi) + \hat{\beta} \sinh(\theta + \phi) \\
\label{Bog4}
&& \hat{b}^\dagger = \hat{\beta}^\dagger \cosh(\theta - \phi) + \hat{\alpha} \sinh(\theta - \phi)
\eea
This map can be thought of as a Bogoliubov transformation
\bea
\label{Bog5}
&& \hat{\alpha}' = \hat{\alpha} \cosh \theta + \hat{\beta}^\dagger \sinh \theta \\
\label{Bog6}
&& \hat{\beta}' = \hat{\beta} \cosh \theta  + \hat{\alpha}^\dagger \sinh \theta
\eea
followed by setting
\bea
\label{Bog7}
&& \hat{a} = \hat{\alpha}' \cosh \phi + \hat{\beta}^{\prime\dagger} \sinh \phi \\
\label{Bog8}
&& \hat{b} = \hat{\beta}' \cosh \phi  - \hat{\alpha}^{\prime\dagger} \sinh \phi
\eea
The Bogoliubov transformation (\ref{Bog5}), (\ref{Bog6}) preserves the canonical commutation relations.
But this is not true of the transformation (\ref{Bog7}), (\ref{Bog8}), due to the relative $-$ sign which appears
in (\ref{Bog8}).  Rather the combined map leads to bulk commutators
\bea
&& [\hat{a},\hat{a}^\dagger] = [\hat{b},\hat{b}^\dagger] = 1 \\
\label{commutator}
&& [\hat{a}^\dagger,\hat{b}^\dagger] = [\hat{b},\hat{a}] = \sinh 2\phi
\eea
with all other commutators vanishing.  Note that $\theta$ drops out of
the commutation relations.  This is expected since $\theta$ in
(\ref{Bog5}), (\ref{Bog6}) parametrizes a Bogoliubov transformation
between the bulk and boundary degrees of freedom, which by definition
is a transformation that preserves the commutators.  Having non-zero
$\phi$, on the other hand, leads to a non-zero commutator between
$\hat{a}$ and $\hat{b}$.  We think of this as representing a bulk
commutator which is non-zero at spacelike separation.  This is a
drastic modification to local field theory -- a risky game to play --
and it's not clear whether a consistent theory can be constructed
along these lines.  But let's proceed, and explore the connection
between non-commutativity and entanglement.

One can start from the microscopic vacuum
\be
\hat{\alpha} \vert 0,0 \rangle = \hat{\beta} \vert 0,0 \rangle = 0
\ee
and build a Fock space
\be
\vert n_\alpha,n_\beta \rangle = {1 \over \sqrt{n_\alpha! n_\beta!}} (\hat{\alpha}^\dagger)^{n_\alpha} (\hat{\beta}^\dagger)^{n_\beta} \vert 0,0 \rangle
\ee
If one only acts on the vacuum with operators of type $a$ one never
notices the non-commutativity (likewise for type $b$).  But suppose
the Hawking particles shown in Fig.\ \ref{figure} correspond to a
two-particle state (which we haven't bothered to normalize)
\bea
\label{ab}
\vert \psi \rangle & \sim & \hat{a}^\dagger \hat{b}^\dagger \vert 0,0 \rangle \\
\nonumber
& \sim & \sinh(\theta + \phi) \vert 0,0 \rangle + \cosh(\theta + \phi) \vert 1,1 \rangle
\eea
In general this state is entangled.  To see this we split the Hilbert
space into $\alpha$ and $\beta$ oscillators, ${\cal H} = {\cal H}_\alpha
\times {\cal H}_\beta$.  The choice of splitting is somewhat
arbitrary, and leads to a freedom that we discuss in more detail
below.  Given the splitting, we construct the density matrix
$\hat{\rho} = \vert \psi \rangle \langle \psi \vert$ and trace over
${\cal H}_\beta$ to obtain the reduced density matrix for particle
$a$.\footnote{A word on notation: the density matrix we are
constructing depends on both our choice of state $\hat{a}^\dagger
\hat{b}^\dagger \vert 0,0 \rangle$, and on our splitting of the
Hilbert space ${\cal H}_\alpha \times {\cal H}_\beta$.  The notation
$\hat{\rho}_a$ emphasizes the former over the latter.}
\be
\hat{\rho}_a = {}_\beta \langle 0 \vert \hat{\rho} \vert 0 \rangle_\beta + {}_\beta \langle 1 \vert \hat{\rho} \vert 1 \rangle_\beta
\ee
Properly normalized, this procedure gives
\be
\hat{\rho}_a = {1 \over 1 + \xi_+^2} \Big( \xi_+^2 \vert 0 \rangle \langle 0 \vert + \vert 1 \rangle \langle 1 \vert \Big)
\ee
where $\xi_+$ is defined by $\xi_+ = \tanh(\theta + \phi)$.  The associated entropy is
\be
\label{Sa}
S_a = - {\rm Tr} \, \hat{\rho}_a \log \hat{\rho}_a = - {\xi_+^2 \over 1 + \xi_+^2} \log \xi_+^2 + \log \left(1 + \xi_+^2\right)
\ee
The mutual information between $a$ and $b$ is $I_{ab} = S_a + S_b - S_{ab}$.  But $S_b = S_a$, while the combined system is in a pure
state with $S_{ab} = 0$.  So the mutual information is simply twice the result (\ref{Sa}),
\be
\label{Iab+}
I_{ab} = - {2 \xi_+^2 \over 1 + \xi_+^2} \log \xi_+^2 + 2 \log \left(1 + \xi_+^2\right)
\ee
Of course this result depends on how we decide to split the Hilbert
space.  In other words, it depends on what we decide to trace over in
constructing $\hat{\rho}_a$.  But the freedom to choose a splitting
can be absorbed into a shift of the Bogoliubov parameter $\theta$.
More precisely $\theta$ parametrizes the freedom to split the Hilbert
space into ${\cal H}_{\alpha'} \times {\cal H}_{\beta'}$, where
$\alpha'$ and $\beta'$ are the independent oscillators defined in
(\ref{Bog5}), (\ref{Bog6}).\footnote{Note that standard field theory
does not have this freedom.  In standard field theory operators
commute at spacelike separation, so one can unambiguously associate
a factor of the Hilbert space with any given spatial region.  Then
for a given region there is no freedom in deciding what to trace
over when computing entropy of entanglement.}

One can use this freedom to set $\theta + \phi = 0$, which makes the mutual information in the state (\ref{ab}) vanish.  But
even if the mutual information in this particular state vanishes, there will still be other states that carry mutual information.  For
example the state
\be
\vert \psi \rangle = \hat{b}^\dagger \hat{a}^\dagger \vert 0,0 \rangle
\ee
has mutual information which can be obtained from (\ref{Iab+}) by replacing $\xi_+ \rightarrow \xi_- \equiv \tanh(\theta - \phi)$, namely
\be
\label{Iab-}
I_{ab} = - {2 \xi_-^2 \over 1 + \xi_-^2} \log \xi_-^2 + 2 \log \left(1 + \xi_-^2\right)
\ee
In attempting to make both (\ref{Iab+}) and (\ref{Iab-}) small the best one can do is set $\theta = 0$.  Then (\ref{Iab+}) and (\ref{Iab-}) are equal,
and at leading order for small $\phi$ the mutual information in either of the states $\hat{a}^\dagger \hat{b}^\dagger \vert 0 \rangle$ or
$\hat{b}^\dagger \hat{a}^\dagger \vert 0 \rangle$ is
\be
\label{FinalIab}
I_{ab} \approx 2 \phi^2(1 - \log \phi^2)
\ee

This toy model suggests that two operators which have a commutator
that is ${\cal O}(\phi)$ as in (\ref{commutator}), typically produce
entangled states with a mutual information that is ${\cal O}(\phi^2)$
as in (\ref{FinalIab}).  Since we know how much mutual information is
present in Hawking radiation, we can estimate how big commutators must
be at spacelike separation.  Our results suggest that two field
operators should have a commutator of order $e^{-S/2}$ in the presence
of a black hole, in order to account for the mutual information
$I_{ab} \sim e^{-S}$ carried by two successive Hawking particles.
(More precisely, we have in mind that matrix elements of the
commutator in a typical state of the black hole plus Hawking radiation
should be of order $e^{-S/2}$.)  In the context of AdS/CFT black hole
entropy is ${\cal O}(N^2)$, so this effect is non-perturbatively small
in the $1/N$ expansion of the CFT.

\bigskip
\goodbreak
\centerline{\bf Acknowledgements}
\noindent
We are grateful to Gilad Lifschytz for valuable discussions and for
comments on the manuscript.  DK thanks the YITP for hospitality during
this work.  NI is supported in part by JSPS KAKENHI Grant Number
25800143.  DK is supported by U.S.\ National Science Foundation grant
PHY-1125915 and by grants from PSC-CUNY.


\providecommand{\href}[2]{#2}\begingroup\raggedright\endgroup

\end{document}